\documentclass{aa}  

\usepackage{graphicx}
\usepackage{txfonts}
\usepackage{here}
\begin{document} 

   \title{Condensation of cometary silicate dust using an induction thermal plasma system}

   \subtitle{I. Enstatite and CI chondritic composition}

   \author{T. H. Kim
          \inst{1}
          \and A. Takigawa\inst{2}
          \and A. Tsuchiyama\inst{3, 4, 5}
          \and J. Matsuno\inst{3}
          \and S. Enju\inst{6}
          \and H. Kawano\inst{7} 
          \and H. Komaki\inst{8}
 }

   \institute{Department of Chemical Engineering, Wonkwang University,
             460 Iksan-daero, Iksan 54538, Republic of Korea\\
              \email{taehee928@wku.ac.kr}
         \and
             Department of Earth and Planetary Science, The University of Tokyo,
             Tokyo 113-0033, Japan\\
             \email{takigawa@eps.s.u-tokyo.ac.jp}
                     \and Research Organization of Science and Technology, Ritsumeikan University, 
             Shiga 525-8577, Japan
                              \and CAS Key Laboratory of Mineralogy and Metallogeny/Guangdong Provincial Key Laboratory of Mineral Physics and Materials, Guangzhou Institute of Geochemistry, Chinese Academy of Sciences (CAS), 
             Guangzhou 510640, China
                                       \and CAS Center for Excellence in Deep Earth Science, 
             Guangzhou 510640, China
             \and The Kyoto University Museum, Kyoto University, 
             Kyoto 606-8502, Japan
             \and Division of Earth and Planetary Sciences, Kyoto University, 
             Kyoto 606-8502, Japan
             \and JEOL Ltd., 
             Tokyo 196-8558, Japan             }


 \abstract{Glass with embedded metal and sulfides (GEMS) is a major component of chondritic porous interplanetary dust particles.
                    Although GEMS is one of the most primitive components in the Solar System, its formation process and conditions have not been constrained.

                    We performed condensation experiments of gases in the system of Mg--Si--O (MgSiO$_3$ composition) and of the S-free CI chondritic composition (Si--Mg--Fe--Na--Al--Ca--Ni--O system) in induction thermal plasma equipment. Amorphous Mg-silicate particles condensed in the experiments of the Mg--Si--O system, and their grain size distribution depended on the experimental conditions (mainly partial pressure of SiO).
                    In the CI chondritic composition experiments, irregularly shaped amorphous silicate particles of less than a few hundred nanometers embedded with multiple Fe--Ni nanoparticles of $\leq$20 nm were successfully synthesized.
                    These characteristics are very similar to those of GEMS, except for the presence of FeSi instead of sulfide grains.
                    We propose that the condensation of amorphous silicate grains smaller than a few tens of nanometers and  with metallic cores, followed by coagulation, could be the precursor material that forms GEMS prior to sulfidation. }
 

 
   

   \keywords{methods: laboratory -- circum-stellar matter -- infrared: ISM: lines and bands -- infrared: stars
               }

   \maketitle
%

\section{Introduction}

Derived from comets, chondritic porous interplanetary dust particles (CP-IDPs) are one of the most primitive materials in the Solar System \citep{RN2528}. Glass with embedded metal and sulfides (GEMS) are amorphous silicate grains of \textasciitilde100 nm in size with embedded FeNi metals and Fe-rich sulfide nanoparticles \citep{RN305}. GEMS are a major component of CP-IDPs and have also been observed in Antarctic micrometeorites \citep{RN2573} GEMS-like material composed of amorphous silicate grains containing nanometals and/or sulfides has been observed in the matrices of primitive carbonaceous chondrite meteorites that have not undergone extensive aqueous alteration \citep{RN2567, RN2572, RN2569}. The difference between the primitive carbonaceous chondrites and GEMS is that GEMS-like material in the matrix of the primitive carbonaceous chondrites is composed of relatively Fe-rich amorphous silicate with Fe sulfides, and no FeNi metals have been observed except for in one meteorite \citep{RN2572}. Therefore, understanding the formation of GEMS and GEMS-like materials and their relationship could be the key to understanding the processes in the early Solar System.

The GEMS in CP-IDPs are either remnants of amorphous silicate dust in the interstellar medium (ISM) or non-equilibrium condensates from the early solar nebula gas. \citet{RN1256} argued that GEMS are survivors of circumstellar dust amorphized in the ISM based on the observation of olivine crystal pseudomorphs in GEMS. This is also supported by the similarity of the infrared (IR) spectra to those of silicate dust observed in molecular clouds and around evolved stars \citep{RN2402}. Conversely, \citet{RN304} proposed that most GEMS were formed by condensation in early solar nebulae as late-stage non-equilibrium condensates based on a low abundance of presolar GEMS grains and the wide range of chemical compositions of GEMS. The formation process and the environment of GEMS remain controversial \citep{RN2558, RN2564}. 

  \begin{figure*} [h]
   \centering
  \includegraphics[width=\hsize]{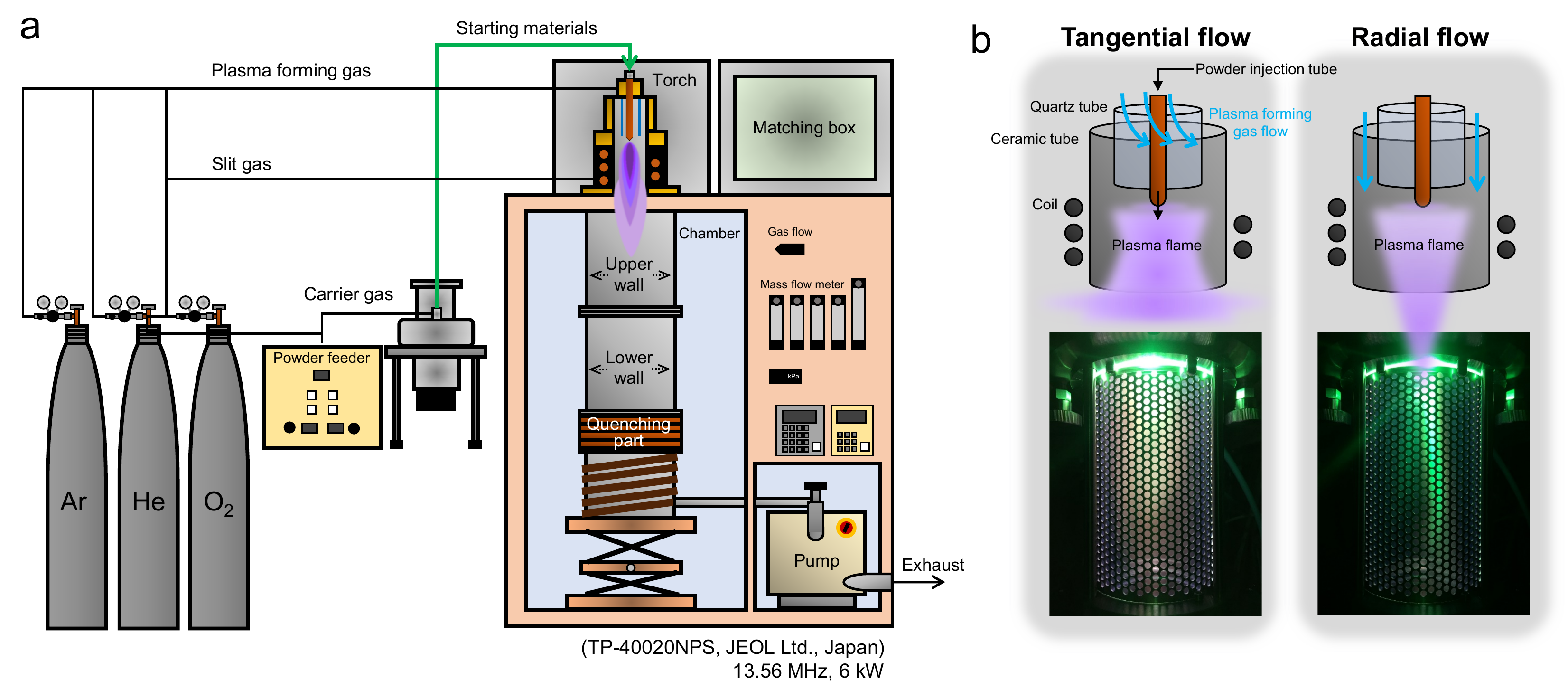}  
   \caption[]{\label{fig:fig_1} %
     Schematic of ITP system for the reproduction of cosmic dust simulants (a) and illustrations and pictures of injection types of plasma-forming gas (b). Tangential flow is a swirl gas flow inside the quartz tube, and radial flow is an axial gas flow between quartz and ceramic tubes.}
 \end{figure*}

\citet{RN2570} succeeded in reproducing GEMS-like material composed of amorphous silicate embedded with metallic Fe--Ni nanoparticles from gases of the GEMS-mean composition (Mg--Fe--Si--Ni--Al--Ca--O) in an induction thermal plasma (ITP) system operated at 30 kW. The characteristics of the condensates closest to those of the GEMS texture (GM7-75) were that (1) condensed amorphous silicate grains were spheres of \textasciitilde45 nm in size, and (2) \textasciitilde10\% of the condensed amorphous silicate grains contain the inclusion of multiple Fe--Ni nanometals with a median size of 16 nm that exhibit a multi-metal glass structure, and the other condensates include amorphous silicates with a single metal core (core-shell structure) and those without metal cores.

Induction thermal plasma systems provide high-temperature plasma for vaporizing refractory materials and high cooling rate conditions to obtain amorphous and/or metastable nanoparticles. A new ITP system (TP-40020NPS, JEOL Ltd.) of 6 kW was installed to study the formation processes of various cosmic dust simulants \citep{RN2524}.

In the present study, we perform condensation experiments using a simple system of MgO-SiO$_2$ and a sulfur-free CI chondritic composition to reproduce GEMS-like grains from a gas. The condensates synthesized in the ITP systems can be compared with the circumstellar dust formation conditions by scaling the experimental conditions in the ITP systems \citep{RN2570, RN2524}. The condensation conditions of GEMS compared with those of the present experiments are discussed based on nucleation and growth theory from vapor \citep{RN2087, RN1581}.

\section{Methods}

 \begin{table*}[h]
        \caption{Experimental conditions and mass of run products.}\label{table:1}
        \centering
        \begin{tabular}{c c c c c c c c c c c}     
                \hline\hline
                \noalign{\smallskip}       
            $^1$Exp.& Plasma & Plasma & Pressure & $^2$Feeding & \multicolumn{4}{c}{Weight of run product (mg)} & $(Pf)^{5/6}$\\
                no. & forming gas & flame & (KPa) & rate & Upper & Lower & Quenching &Total & value\\
                & (L/min) & flow & & (mg/min) & wall & wall & part & & (kPa$\cdot$mg/min)\\
                \noalign{\smallskip}
                \hline                    
                \noalign{\smallskip}
                & Ar 30 & & & & & & & & \\
                $^3$MS-1 & He 2.5 & Tangential & 33 & 61 & 797.3 & 212.4 & 203.5 & 1213.2 & 564\\
                & O$_2$ 0.5 & & & & & & & &\\
                \noalign{\smallskip}            
                & Ar 30 & & & & & & & & \\
        MS-2 & He 2.5 & Tangential & 30 & 68 & 1092 & 171 & 89 & 1352 & 570\\
        & O$_2$ 0.5 & & & & & & & &\\           
                \noalign{\smallskip}
                & Ar 30 & & & & & & & & \\
                MS-3 & He 2.5 & Tangential & 30 & 57 & 907.8 & 115 & 113.8 & 1136.6 & 493\\
                & O$_2$ 0.5 & & & & & & & &\\           
                \noalign{\smallskip}
                & Ar 30 & & & & & & & & \\
                MS-4 & He 2.5 & Tangential & 50 & 18 & 322.3 & 40.4 & 3 & 365.7 & 293\\
                & O$_2$ 0.5 & & & & & & & &\\           
                \noalign{\smallskip}
                & Ar 30 & & & & & & & & \\
                MS-5 & He 2.5 & Tangential & 70 & 39 & 648.4 & 112.3 & 20.7 & 781.4 & 731\\
                & O$_2$ 0.5 & & & & & & & &\\           
                \noalign{\smallskip}
                & Ar 30 & & & & & & & & \\
                MS-6 & He 2.5 & Radial & 72 & 35 & 174.8 & 414.8 & 108 & 697.6 & 681\\
                & O$_2$ 0.5 & & & & & & & &\\           
                \noalign{\smallskip}
                \hline  
                \noalign{\smallskip}     
                CI-1 & Ar 30 & Tangential & 30 & 174 & \multicolumn{2}{c}{3280}& 198 & 3478 & 1253 \\
                & He 2.5 & & & & \multicolumn{2}{c}{(mixed)}& & & \\                 
                \noalign{\smallskip}
                CI-2 & Ar 30 & Tangential & 70 & 71 & 1100.3 & 160.8 & 150 & 1411.1 & 1197 \\
                & He 5.0 & & & & & & & & \\                  
                \noalign{\smallskip}
                CI-3 & Ar 30 & Radial & 70 & 3 & 14.1 & 46.9 & 8.7 & 69.7 & 98 \\
                & He 5.0 & & & & & & & & \\                  
                \noalign{\smallskip}
                \hline          
        \end{tabular}
        \tablefoot{\\
        $^1$The MS-1 to MS-6 experiments were performed in system MgO-SiO$_2$, and the CI-1 to CI-3 experiments were performed in system Si--Mg--Fe--Na--Al--Ca--Ni--O.\\
                $^2$Starting materials were injected with argon carrier gas at 3 L/min.\\
                $^3$Oxygen slit gas (O$_2$) was additionally injected at 20 L/min.\\ 
        }
   \end{table*}

\subsection{Experimental methods}
A schematic illustration of the ITP system is shown in Fig.1a.
The system consists of a plasma torch, a chamber, a quenching part, and a powder feeder for injection of the starting materials. The plasma torch, the wall of the chamber, and the quenching part are cooled by running water. The ITP is generated and sustained by the induction heating mechanism \citep{RN2565}. The heat energy generated via the Joule heating process is sufficient for continuous dissociation and ionization of the injected gas and thus for sustaining the plasma. The generated high-temperature (\textasciitilde15,000 K) plasma flame provides effective vaporization of the injected starting materials, and a steep temperature gradient (10$^4$--10$^5$ K/s) enables condensation of the nanoparticles.

The input power was fixed at 6 kW. To generate a stable plasma flame by inductively coupled discharge, Ar gas was used as the main plasma-forming gas owing to its high electrical conductivity. He gas was added to improve the heat transfer to the starting material because of its high thermal conductivity, and O$_2$ gas was added to adjust the redox conditions in the system.\\
Plasma-forming gases in conventional ITP systems are injected both inside and outside of the quartz tube at the torch. In contrast, in the ITP system employed in this study, the direction of the plasma-forming gas was either inside or outside of the quartz tube to form a distinctly different high-temperature pattern in the plasma flame. We applied one of the two injecting directions of the plasma forming-gases, called tangential and radial flows, in each experiment (Fig. 1b). In the case of the tangential gas flow, the plasma-forming gas was injected to swirl inside the quartz tube and expand broadly upon exiting the torch (Fig. 1b, left). The radial plasma-forming gas was introduced vertically outside the quartz to generate a long plasma column with vertical gas injection (Fig. 1b, right).

We performed two sets of experiments, which we refer to as MS and CI experiments, with starting materials composed of Mg/Si = 1 and of CI chondritic composition, respectively. The experimental conditions are detailed in Table 1.

In the MS experiments, we performed six runs (MS-1 to MS-6) in the system of MgO--SiO$_2$, which is one of the simplest systems that mimic the solid materials in the Solar System to obtain the optimal conditions for evaporation and condensation by changing the operating conditions of the ITP system. We used a mixture of micron-sized powders of periclase (MgO, 99\%, <30~$\mu$m; Wako Pure Chemical Industries Ltd., Japan) and quartz (SiO$_2$, 99.9\%, 4~$\mu$m; Kojundo Chemical Lab. Co., Ltd., Japan) with an atomic ratio of 1:1, or enstatite composition. The following parameters were examined: application of an additional slit gas (MS-1 and MS-2), reproducibility (MS-2 and MS-3), reactor pressure (MS-3, MS-4, and MS-5), and injection direction of the plasma-forming gas (MS-5 and MS-6). As the plasma-forming gas, we used a mixed gas of Ar, He, and O$_2$ at 30, 2.5, and 0.5 L/min, respectively. To prevent the formation of metal nanoparticles, O$_2$ gas was added.

The CI experiments (CI-1, CI-2, and CI-3) were performed in the Si--Mg--Fe--Na--Al--Ca--Ni--O system. The elemental ratio of the cations of the starting material was prepared as the CI chondritic composition \citep{RN2568} (Table 2). For the starting material, we prepared a mixture of micron-sized oxides and metallic powder reagents containing Si (>99.9\%, 5~$\mu$m; Kojundo Chemical Lab. Co., Ltd., Japan), SiO$_2$, MgO, Fe (>99.9\%, 3--5~$\mu$m, Kojundo Chemical Lab. Co., Ltd., Japan), $\gamma$-Al$_2$O$_3$ (>99.9\%, 4~$\mu$m; Kojundo Chemical Lab. Co., Ltd., Japan), CaO (>99.95\%, 4~$\mu$m; Strem Chemicals Inc., USA), anhydrous Na$_2$SiO$_3$ (<100~$\mu$m; Nacalai Tesque Inc., Japan), and Ni (>99.9\%, 2--3~$\mu$m; Kojundo Chemical Lab. Co., Ltd., Japan). We adopted a ratio of Si/(SiO$_2$ + Si) of 0.3 for controlling the O$_2$ content in the starting material to form both amorphous silicates and Fe--Ni metals \citep{RN2570}. The mixture was ground to particles of ~4~$\mu$m.

The starting material was injected into the plasma flame with Ar carrier gas at 3 L/min for 20 min in all runs. The same plasma-controlling conditions of MS-2, MS-5, and MS-6 were adopted for CI-1, CI-2, and CI-3, respectively. Ar and He gases were used as the plasma-forming gas. The flow rate of the He was increased to 5 L/min to improve the heat transfer in CI-2 and CI-3. The pressure was controlled at 30 kPa and 70 kPa in CI-1 and CI-2, respectively, and different injection patterns of the plasma-forming gas (i.e., tangential and radial) were applied in CI-2 and CI-3 (Table 1).

As the thermal plasma torch is located above the chamber (Fig. 1a), the gas temperature is highest at the top of the chamber and steeply decreases due to a high fluid velocity, resulting in the condensation of small particles. Most of the condensates are transported along with the plasma gas and coagulate on the walls of the chamber; some of the condensates, together with evaporation residues, fall down onto the quenching part, which is directly below the plasma torch, due to gravity. The run products were separately collected at the upper and lower walls of the chamber (Fig. 1a) because of the anisotropic temperature gradients in the chamber \citep{RN2577}. The exception is run CI-1, in which the condensates at the upper and lower walls of chambers were collected as a mixture. We also collected samples at the quenching part to obtain the total mass of the products. The feeding rate of the starting material is calculated from the total mass of the produced powder and the experimental duration (Table 1).

\subsection{Analytical methods}
The phases of the starting materials and run products were examined using powder X-ray diffraction (XRD; SmartLab, Rigaku, Kyoto University) and Fourier transform infrared (FT-IR) spectroscopy (MFT-680; JASCO, Kyoto University). X-ray diffraction with Ni-filtered CuK$\alpha$ radiation ($\lambda$ = 1.5418 \AA) was performed at an accelerating voltage and tube current of XRD of 40 kV and 40 mA, respectively. The run products were mounted on a reflection-free sample holder of a Si single crystal for detecting amorphous materials. The scan range was 4$^{\circ}$--95$^{\circ}$~2$\theta$; the scan step was 0.02$^{\circ}$ count; and the scan speed was 0.3$^{\circ}$/min. For FT-IR analysis, each sample and KBr powder were mixed at a mass ratio of 1:200 and were pressed to form a 10 mm $\phi$ pellet. The absorbance of the pellets was measured at a wavelength range 1.6--25~$\mu$m. For observation using a field-emission scanning electron microscope (FE-SEM; JSM-7001F, JEOL, Kyoto University) with an energy dispersive X-ray spectrometer (EDS; XMax-150, Oxford, Kyoto University), each run product was embedded into resin and was polished. A small amount of each run product was put on a copper mesh transmission electron microscope (TEM) grid coated by a carbon film and was observed using FE-TEMs (JEM-2100F, JEOL, Kyoto University; JEM-2800, JEOL, University of Tokyo) with energy dispersive X-ray spectroscopy EDS (JED-2300T, JEOL, Kyoto University; X-Max50, Oxford, University of Tokyo) at 200 kV accelerating voltage. Scanning TEM (STEM) images and STEM-EDS maps were obtained to determine the textures and phases of each grain in the condensates. In addition, STEM tomography was performed to obtain three-dimensional (3D) textures of the amorphous silicate grains with embedded metallic nanoparticles synthesized in CI-2. For STEM-EDS tomography, the nanoparticles were mounted on carbon thin film attached to copper mesh, and TEMography (System In Frontier Inc., Japan) was used for acquisitions and analysis. Dark field (DF) STEM and EDS images were acquired for every 8$^{\circ}$ of the sample stage rotation between $-63.5^{\circ}$~and +62$^{\circ}$~for 20 min, and 17 to 18 images in total were used for the tilt series. Each image contained 256 $\times$ 256 pixels at a pixel size of 2--4 nm. After semi-manually estimating the rotation axis, we reconstructed the tilt series using the simultaneous iterative reconstruction technique (SIRT) algorithm and obtained tomographic images. The volume rendering image was drawn by giving an appropriate transmittance to the pixel values of the tomographic image and was observed in 3D.
\section{Results} \label{sec:Results}

\subsection{MS experiments in the MgO-SiO$_2$ system}
Scanning electron microscope and XRD analysis showed that collected samples are dominated by condensates of amorphous Mg silicates and contain crystalline and glassy coarse grains, which are unvaporized residues of the starting materials (Fig. A.1 and A.2). Figure 2 shows the FT-IR spectra of the starting material and run products collected from the upper wall of the chamber. The broad peaks of the products at \textasciitilde9.6~$\mu$m and 18.7~$\mu$m indicate that most of the condensates are amorphous Mg silicates. The peaks are attributed to Si--O stretching and O--Si--O bending modes of the amorphous Mg silicates, respectively. The observed peak at \textasciitilde9.6~$\mu$m is clearly different from that of \textasciitilde9.3~$\mu$m for the amorphous silicates with Mg/Si ratios of 0.7--1 synthesized by a sol-gel method \citep{RN473}, which may be due to different atomic structures of amorphous silicate and also affected by evaporation residues. The peaks of amorphous silica at \textasciitilde9.0~$\mu$m and 20.8~$\mu$m \citep{RN2566} are relatively strong in MS-1. Shoulders at 11.2~$\mu$m attributed to crystalline forsterite \citep{RN2398} also occurred in MS-1 to MS-3. These crystals are not condensates but unvaporized residues of starting materials.

\begin{figure}[hbtp]
        \centering
        \includegraphics[width=\columnwidth]{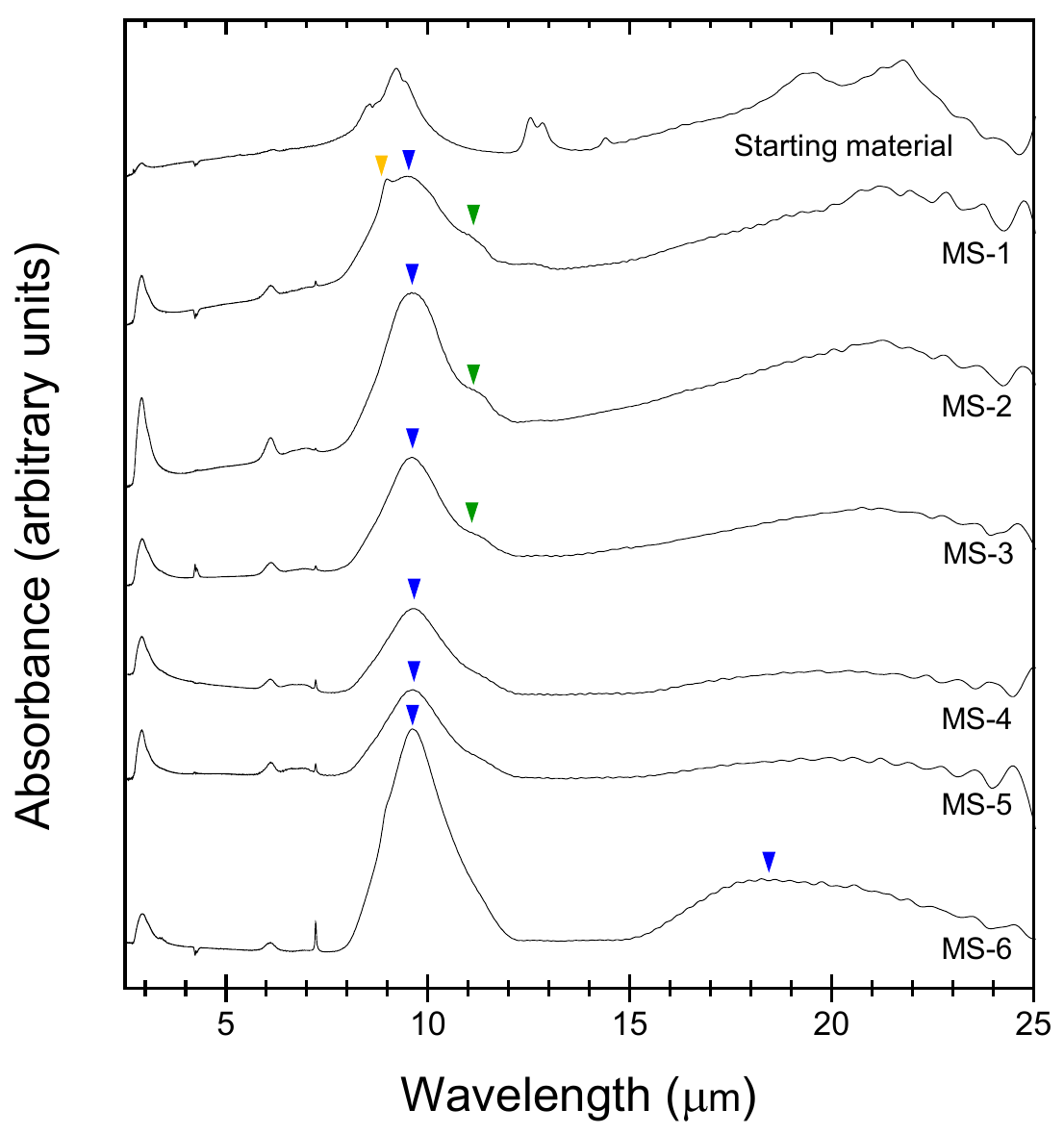}  
        \caption[]{\label{fig:fig_2} %
                FT-IR absorbance of run products collected from the upper wall in MS-1--6. Blue, green, and yellow arrows indicate peaks attributed to amorphous silicate, forsterite, and amorphous silica, respectively.
        }
\end{figure}

The TEM observation showed that fine-grained materials correspond to condensed products composed of nanoparticle aggregates (Fig. 3). Spherical grains of 10--100 nm with clear grain boundaries are dominant in the condensates at the lower wall in MS-1--5 and from the upper and lower walls of the chamber in MS-6. On the other hand, small particles <50 nm obtained from the upper wall in MS-1--5 are nonspherical with rough surfaces and less-clear grain boundaries (Figs. 3a and b). The selected area electron diffraction (SAED) of the nanoparticles indicate amorphous state (Figs. 3c and e); no crystalline phase was detected in the condensates of MS-1--6.

For the experiments using tangential gas flow (MS-1--5), the largest sizes of the nanoparticles are smaller in the upper wall than those in the lower wall (Figs. 3c and d). No clear size difference was noted in the particles from the upper and lower walls of chambers in experiments that used radial gas flow (MS-6; Figs. 3e and f). 

To discuss the grain size difference more quantitatively, we obtained size distributions of the run products using the TEM images of the condensates obtained from the lower wall in MS-1--4 and from the upper and lower walls in MS-5 and MS-6 (Fig. 4). The grain size of the condensates from the upper wall in MS-1--4 was not measured owing to their unclear grain boundaries (Fig. 3a). The grain size distribution also depended on the gas flow direction (Fig. 4a). The condensates from the lower wall of the chamber in MS-1--5 are similar but are clearly smaller than those in MS-6 (Fig. 4a). Figure 4b compares the size distributions of condensates from the upper and lower walls of the chambers in MS-5. The median sizes of particles from the upper and lower wall are 13.4 nm and 29.4 nm, respectively. The condensates from the {lower wall} are twice the size and have a broader size distribution than those of the upper wall. In MS-6, the median particle size from the upper and {lower} chambers is similar, at 32.9 nm and 39.0 nm, respectively, and is larger than those in MS-5 (Fig. 4c). The chemical compositions of nanoparticles in the upper wall are homogeneous, whereas those from the lower wall are clearly heterogeneous and have Mg-rich cores and Si-rich rims (Fig. A.3) This discrepancy could be attributed to the immiscibility gap of silicate melts, as discussed subsequently. The optimal condition was obtained for evaporation and condensation in the MgO-SiO$_2$ system and it was applied to the CI experiments (Appendix. B).

\begin{figure}[hbtp]
        \centering
        \includegraphics[width=\columnwidth]{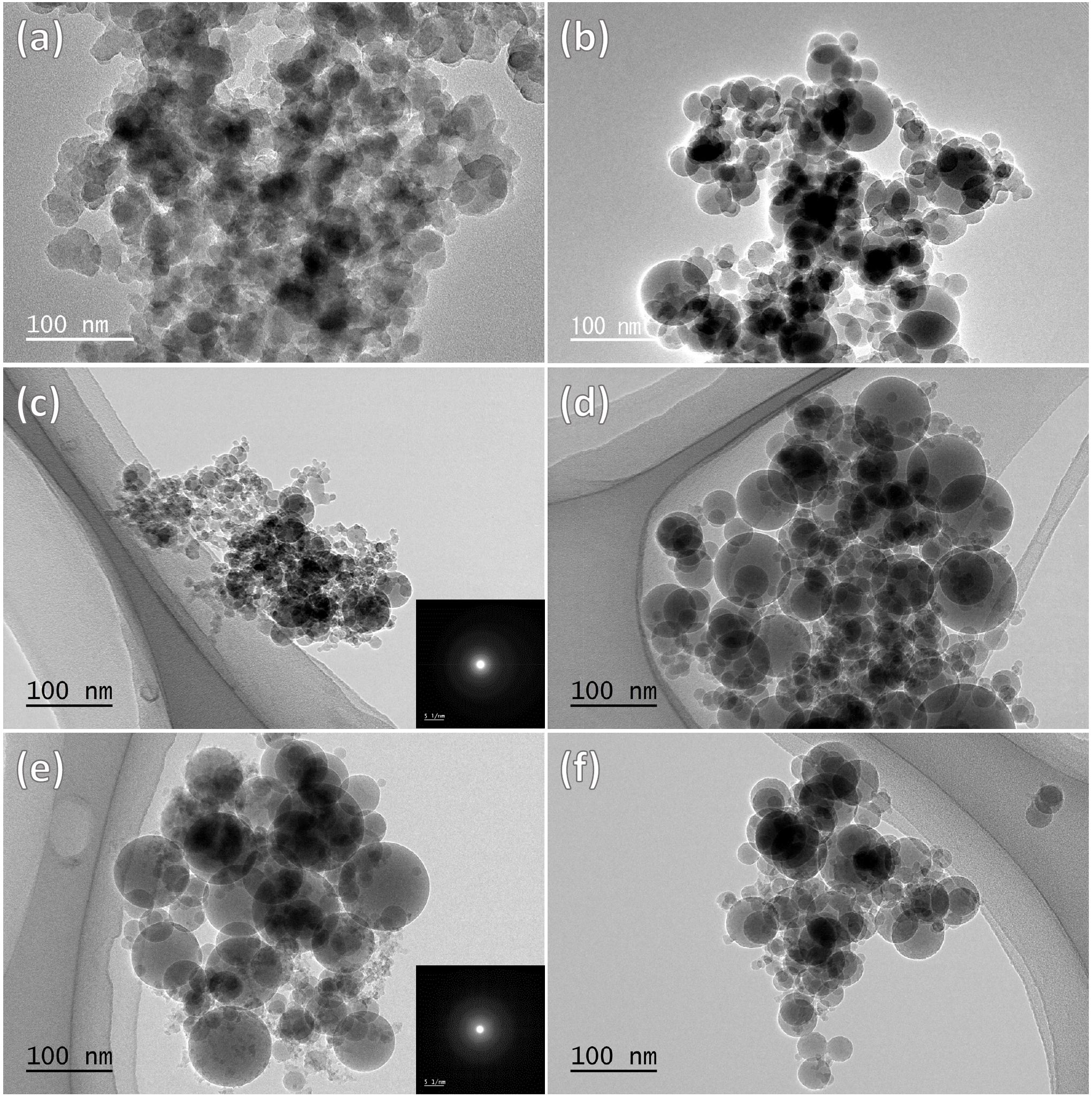}  
        \caption[]{\label{fig:fig_3} %
                TEM-BF images of run products collected from the upper {wall} in MS-1  (a), {the lower wall} in MS-2 (b), the upper (c) and {lower (d) walls} in MS-5, and the upper (e) and {lower walls} (f) in MS-6.
        }
\end{figure}
\begin{figure}[hbtp]
        \centering
        \includegraphics[width=\columnwidth]{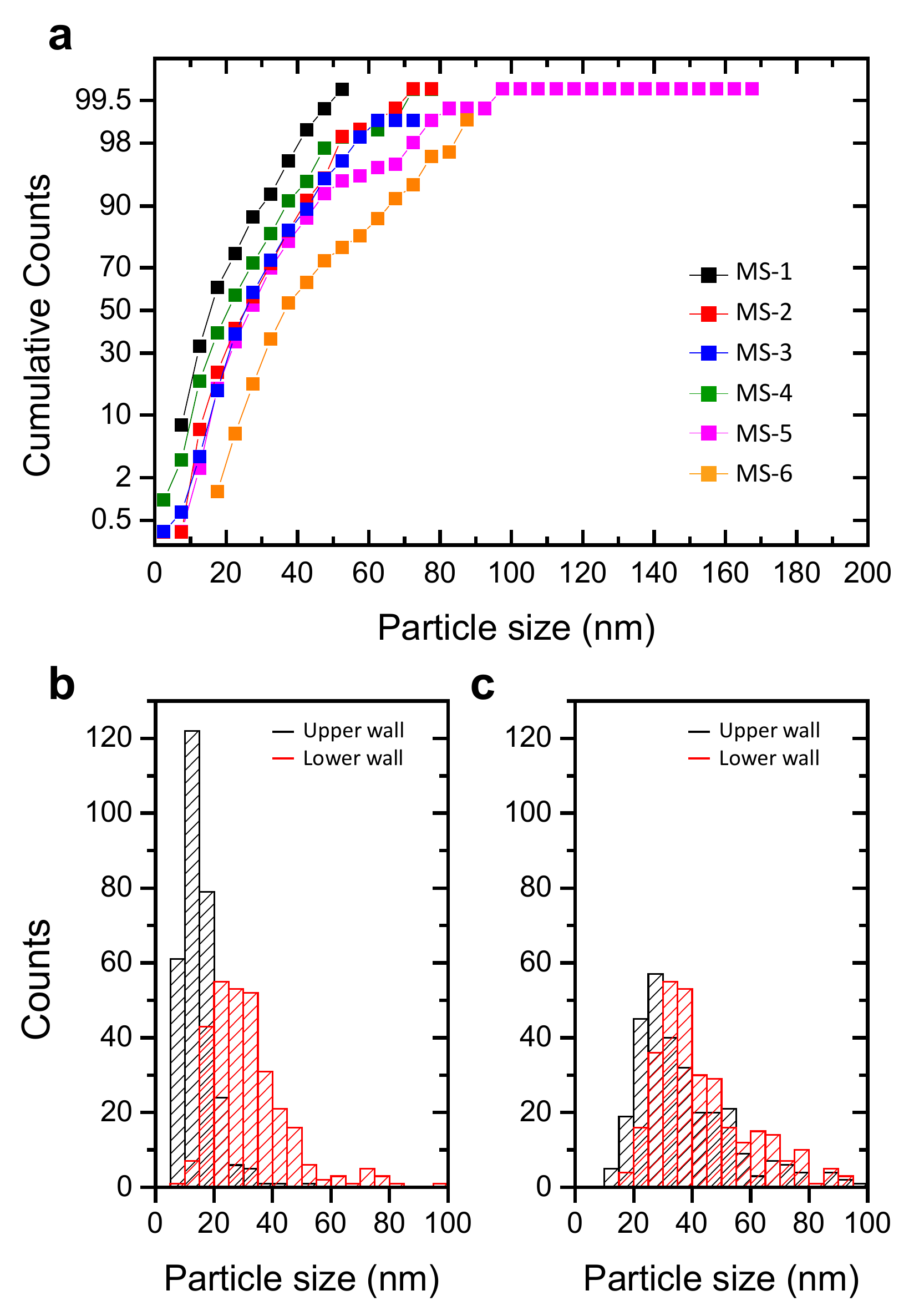}  
        \caption[]{\label{fig:fig_4} %
                Size distribution of run products collected from the {lower wall} in MS-1-6 (a), the upper and {lower walls} in MS-5 (b), and the upper and {lower walls} in MS-6 (c).
        }
\end{figure}

\subsection{CI experiments in the Si--Mg--Fe--Na--Al--Ca--Ni--O system}
\subsubsection{{Bulk analysis}}
{Analysis of SEM, XRD, and FT-IR showed that the condensates are mainly amorphous silicates and Fe-Ni metals and silicides. Unvaporized residues of the starting materials are mainly observed in the quenching part in the CI experiments.

Figure 5 shows the XRD patterns of the starting material and run products in CI-1--3. The run products show halos at 20$^{\circ}$--40$^{\circ}$ with weak peaks from crystals, indicating that most of the condensates are amorphous silicates in the experiments. Samples from the quenching part show strong peaks of unvaporized residues of the starting materials such as periclase, quartz, silicon, corundum, metallic Fe, and lime. Forsterite is the crystallization product from unevaporated melt, as in the MS experiments. For the samples collected from the chamber walls, peaks from unvaporized residues are weak in CI-1 and negligible in the sample in CI-2 and CI-3. The proportion of evaporation residues decreases with increases in the reactor pressure. The run products are mostly composed of condensates in CI-3, where the radial plasma flame pattern was chosen (Fig. 5d). Most of the metallic Fe, Fe$_3$Si (gupeiite), and (Fe, Ni) (kamacite, hereafter FeNi) detected in the XRD patterns are condensation products, as indicated by the TEM observation described subsequently.} The IR spectra of the run products collected have broad peaks at \textasciitilde10.0~$\mu$m in CI-1 and CI-2, whereas the peak is shifted to \textasciitilde9.7~$\mu$m in CI-3 {(Fig. 6)}. These peaks are derived from Si--O stretching vibrations in amorphous Mg(Fe)-silicates. The difference in the peak positions could reflect the Mg/Fe and Mg/Si ratios in the amorphous silicates.

\begin{figure*}[hbtp]
        \centering
        \includegraphics[width=\hsize]{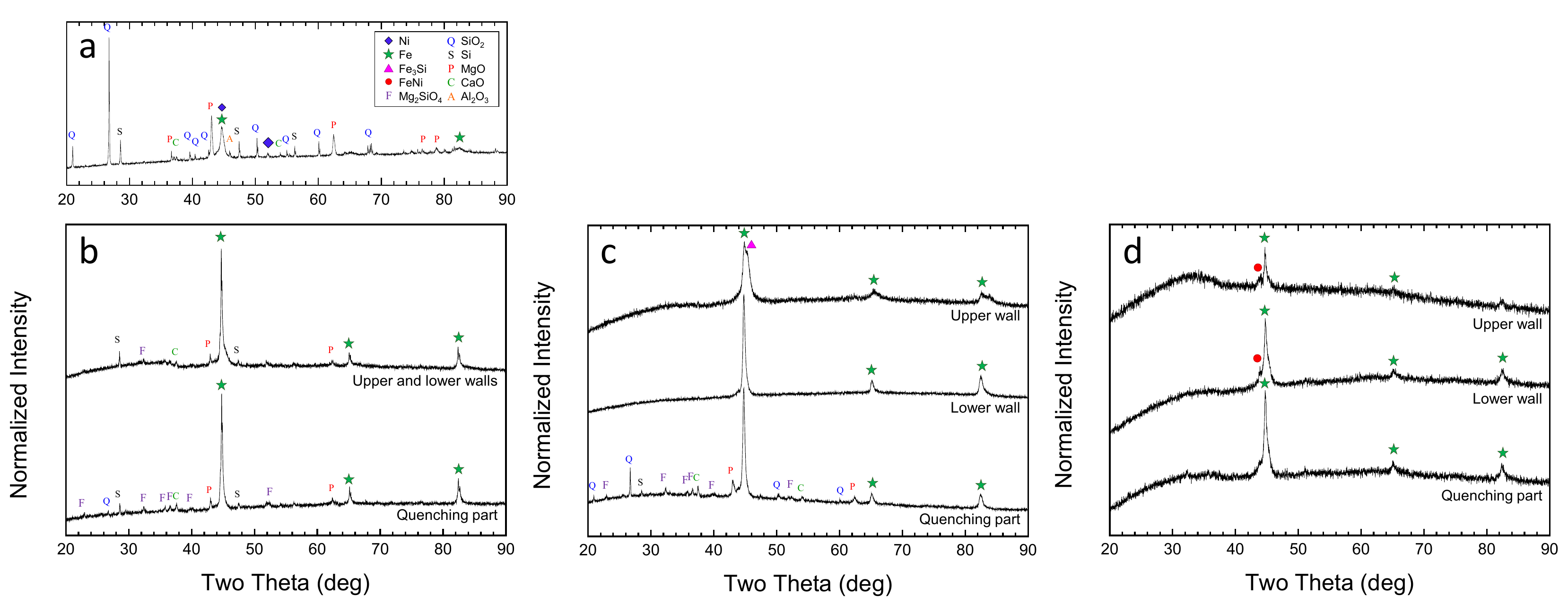}  
        \caption[]{\label{fig:fig_5} %
                XRD patterns of starting material (a) and run products in CI-1 (b), CI-2 (c), and CI-3 (d). }
        
\end{figure*}

\begin{figure}[hbtp]
        \centering
        \includegraphics[width=\columnwidth]{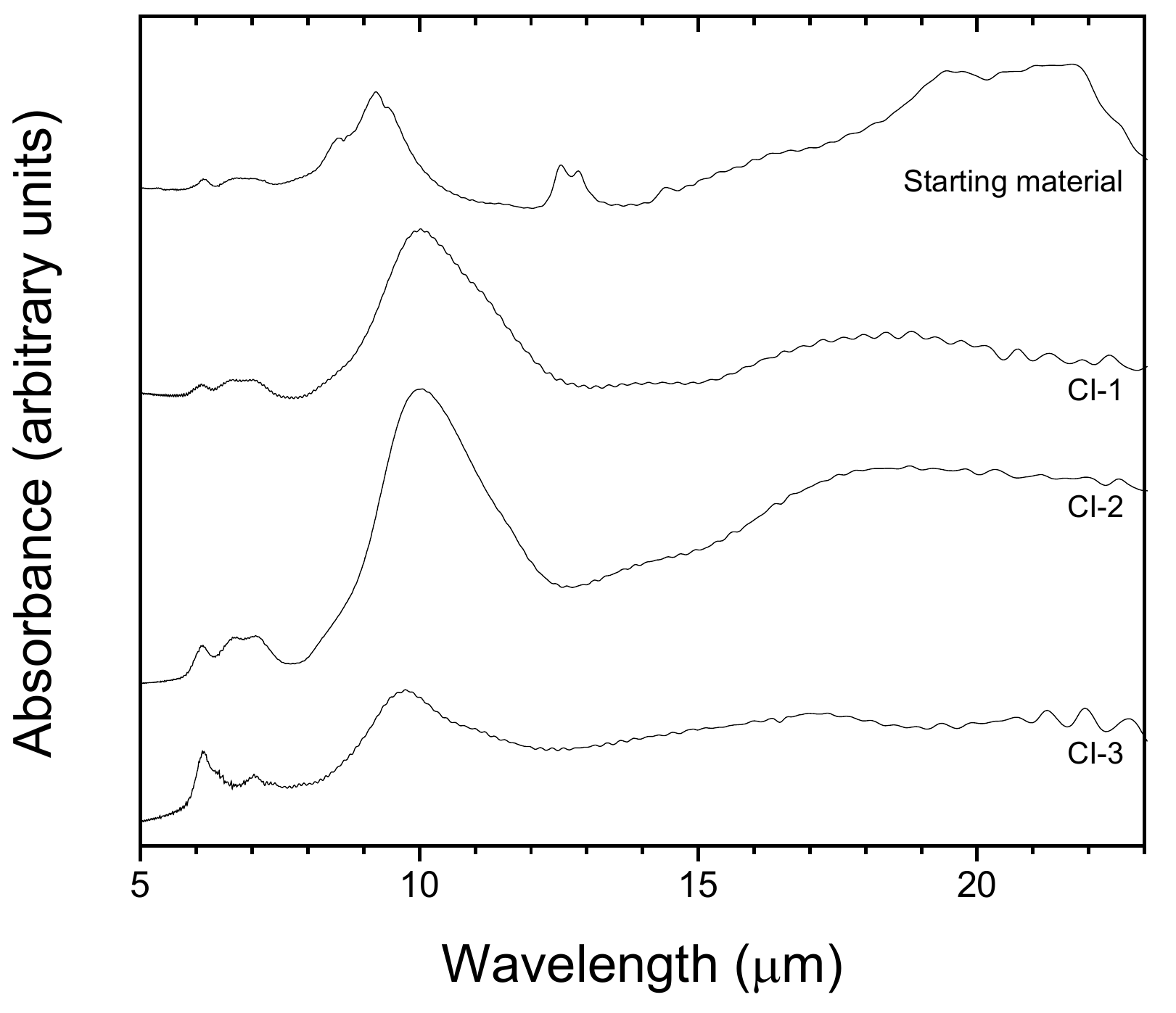}  
        \caption[]{\label{fig:fig_6} %
                FT-IR absorbance of run products (upper wall) in CI-1--3. }
\end{figure}

\subsubsection{Particle analysis}
{Figure 7} shows {bright field (BF) TEM} images of the run products in CI-1--3. High-angle annular dark field (HAADF) STEM images and STEM-EDS maps of the condensates in CI-1 and those from the upper {wall} in CI-2 and CI-3 are shown in {Fig. 8. The results of the TEM observation are summarized in Table 3.}

In the run products of CI-1, in which condensates from the upper and {lower walls} were mixed, amorphous silicates of <100 nm in diameter accompanied by Fe-rich nanoparticles <\textasciitilde20 nm were observed {(Figs. 7a and 8a). Relatively large (>30 nm) amorphous silicate grains are perfectly spherical and show a clear grain boundary.} The nanoinclusions are mainly metallic FeNi and Fe$_3$Si nanoparticles {(Fig. 8a)}, which could correspond to the kamacite and gupeiite observed in the XRD patterns, respectively {(Fig. 5)}. Not more than one nanoparticle is present in each amorphous silicate grain. This texture is similar to that reported in the previous work as a core-shell microstructure \citep{RN2570}. The Fe/(Fe + Mg) of amorphous silicates from the upper {wall} in CI-1 is \textasciitilde0.2 (Table 2).

The amorphous silicates collected from the {lower wall} in CI-2 are larger than those from the upper {wall} and have spherical shapes with clear grain boundaries. No more than one metallic sub-grain is embedded in each amorphous silicate grain from the {lower wall} in CI-2 {(Fig. 7c)} as in the run products in CI-1. For the CI-2 run products collected from the upper {wall}, most amorphous silicate particles greater than tens of nanometers are irregularly shaped with no clear grain boundaries {with multiple metallic inclusions (Fig. 7b). They are aggregates of small <\textasciitilde20 nm amorphous silicate grains with and without a single metallic core. The rough surface and no clear grain boundaries indicate either that nonspherical grains were originally condensed and aggregated, or that spherical nanoparticles condensed and coalesced and then surface atoms diffused to modify the grain shapes. The Fe content in the amorphous silicates is relatively low, at (Fe/(Fe + Mg) = \textasciitilde0.09; Table 2). The embedded nanoparticles are composed of FeNi and Fe$_3$Si; multiple nanoparticles are present in the amorphous silicate grains from the upper wall {(Figs. 7b and 8b)}. The XRD pattern indicates that Fe$_3$Si is dominant over FeNi {(Fig. 5c)}.

In the run products of CI-3, the condensates from the {lower wall} are similar in texture and size to those from the upper {wall (Figs. 7d and e) and those from the upper wall in CI-2 (Fig. 7c)}. The amorphous silicates from the {lower wall} in CI-3 are enriched in Fe (Fe/(Fe + Mg) = \textasciitilde0.3; Table 2), and some grains contain inclusions of mostly FeNi and, to a lesser extent, Fe$_3$Si {(Fig. 8c). The experimental condition may be more oxidized in CI-3 than in CI-2 because considering that the TEM observation showed abundant nanoparticles of FeNi and Fe$_3$Si and Fe-poor amorphous silicate particles in CI-2, whereas metallic nanoparticles were rare, and the amorphous silicate was Fe-rich in CI-3.
 
The chemical compositions of the Mg- and Si-rich domains of amorphous silicates were extracted from the STEM-EDS maps (Table 2). In all run products of CI-1--3, many amorphous silicate grains had Si-rich cores and Mg-rich rims {(Figs. 8a--c)}, as was the case of the MS experiments.

\begin{figure}[hbtp]
        \centering
        \includegraphics[width=\columnwidth]{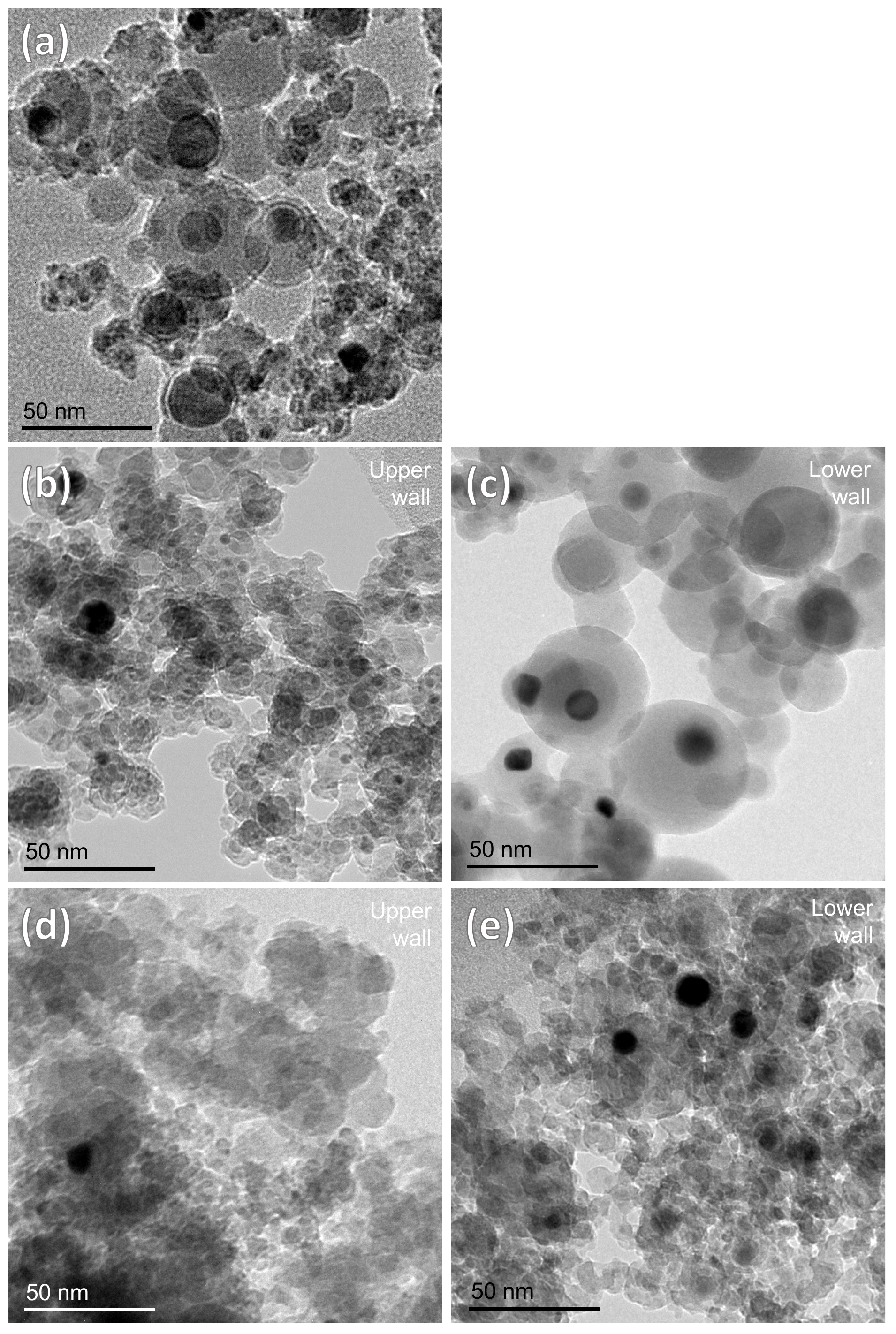}  
        \caption[]{\label{fig:fig_7} %
                TEM-BF images of run products from the upper and {lower walls} in CI-1  (a), the upper (b) and {lower walls (c)} in CI-2, and the upper (d) and {lower walls} (e) in CI-3. }
\end{figure}
\begin{figure*}[hbtp]
        \centering
        \includegraphics[width=\hsize]{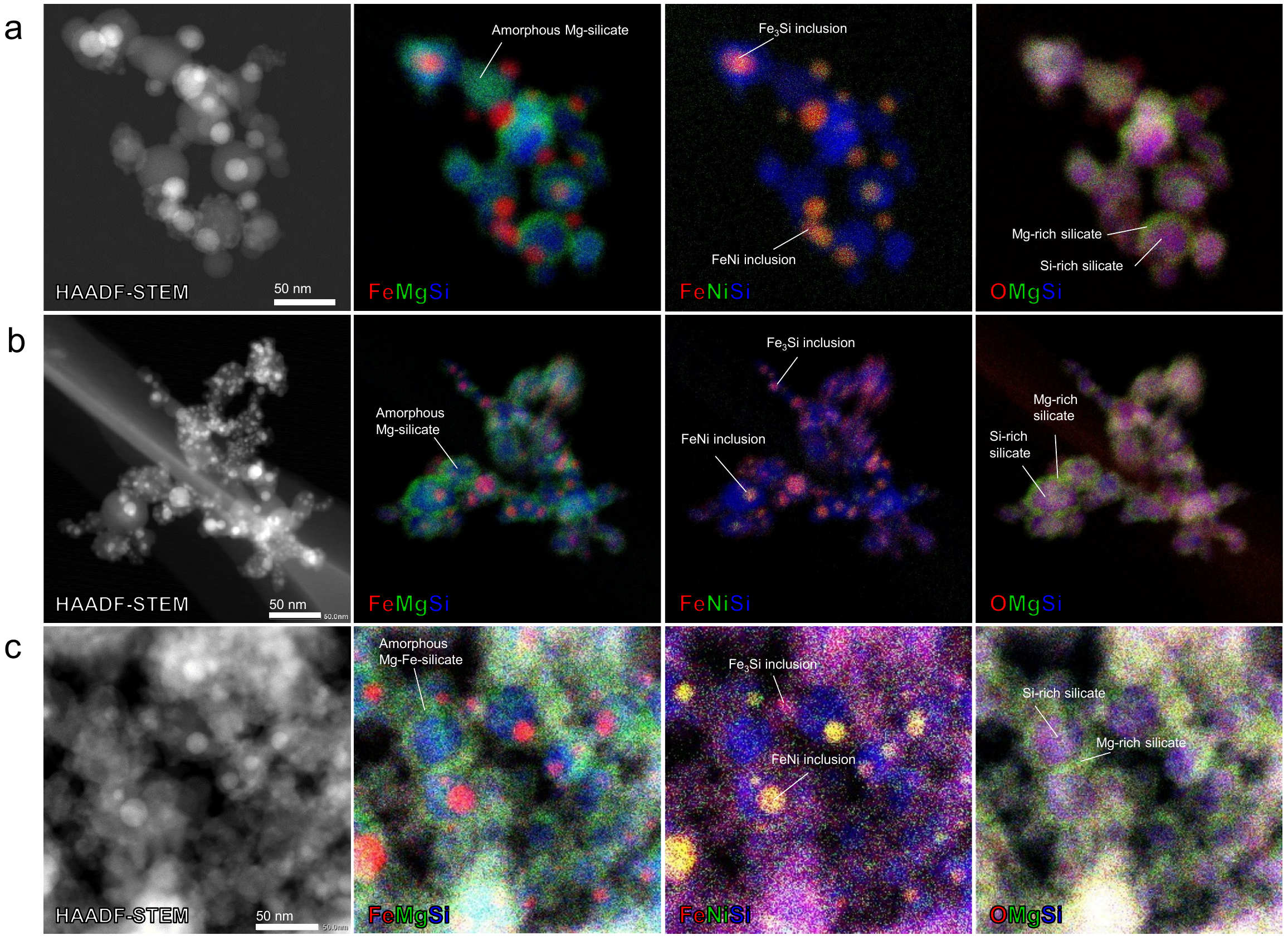}  
        \caption[]{\label{fig:fig_8} %
                HAADF-STEM images and STEM-EDS maps of run products synthesized in CI-1 (a), CI-2 (upper {wall}) (b), and CI-3 (upper {wall}) (c).}
\end{figure*}
\begin{table*}
        \caption{Representative chemical composition of the amorphous silicate condensates determined by STEM-EDS.}             
        \label{table:2}      
        \centering          
        \begin{tabular}{l c c c c c c c c}     
                & & & & & & & (at. \%)\\
                \hline\hline
                \noalign{\smallskip}       
                & Starting & \multicolumn{2}{c}{CI-1} & \multicolumn{2}{c}{CI-2} & \multicolumn{2}{c}{CI-3}\\             
                & material & Mg-rich & Si-rich & Mg-rich & Si-rich & Mg-rich & Si-rich\\ 
                \noalign{\smallskip}
                \hline  
                \noalign{\smallskip}
                Na & 0.06 & 0.01 & 0.00 & 0.00 & 0.00 & 0.07 & 0.02\\
                Mg & 1.03 & 1.52 & 0.31 & 2.46 & 0.69 & 2.17 & 0.61\\
                Al & 0.08 & 0.01 & 0.02 & 0.04 & 0.04 & 0.01 & 0.04\\
                Si & 1    & 1.00 & 1.00 & 1.00 & 1.00 & 1.00 & 1.00\\
                Ca & 0.06 & 0.01 & 0.02 & 0.08 & 0.04 & 0.08 & 0.03\\
                Fe & 0.85 & 0.46 & 0.01 & 0.23 & 0.06 & 0.84 & 0.08\\
                Ni & 0.05 & 0.01 & 0.00 & 0.02 & 0.00 & 0.09 & 0.02\\
                \noalign{\smallskip}
                \hline                  
        \end{tabular}
\end{table*}
\begin{table*}
        \caption{Summary of synthesized GEMS-like materials.}             
        \label{table:3}      
        \centering          
        \begin{tabular}{c c c c c c}     
                \hline\hline
                \noalign{\smallskip}       
                &  &  & CI-1 (mixture of upper and lower) & CI-2 (upper) & CI-3 (upper)\\ 
                \noalign{\smallskip}
                \hline
                \noalign{\smallskip}  
                Amorphous & Size (nm) & & 20--50 & 10--100 & 50--100\\\cline{2-6}
                \noalign{\smallskip}
                silicates & Fe/(Mg+Fe) in Mg-rich domain & & 0.2 & 0.09 & 0.3\\
                \noalign{\smallskip}
                \hline
                \noalign{\smallskip}                  
                & Phase and Size (nm) & FeNi & Common & Rare & Common\\
                &  &  & (10--30) & (<10) & (10--40)\\
                \noalign{\smallskip}  
                Inclusions &  & Fe$_3$Si & Rare & Common & Rare\\
                &  &  & (10--20) & (<15) & (<10)\\\cline{2-6}
                \noalign{\smallskip}     
                & Number density & & {Medium} & High & Low\\
                \noalign{\smallskip}
                \hline
                \noalign{\smallskip}
                Morphology & & & Spherical & Nonspecific & Nonspecific\\
                & & & (core-shell) & (GEMS) & \\
                \noalign{\smallskip}                    
                \hline                  
        \end{tabular}
\end{table*}

{Figure 9} shows BF-STEM and HAADF-STEM images of two typical assemblages of grain collected from the upper {wall} in CI-2. The texture of multiple metal nanoparticles in somewhat irregularly shaped amorphous silicate grains in CI-2 is very similar to GEMS except for the presence of Fe$_3$Si and the absence of FeS \citep{RN304}. The 3D texture of this run product obtained by STEM and EDS tomographies shows that Fe$_3$Si and FeNi nanoparticles were embedded in the amorphous silicate rather than attached to the surface (Fig. 10; see also online movie).

\begin{figure}[hbtp]
        \centering
        \includegraphics[width=\columnwidth]{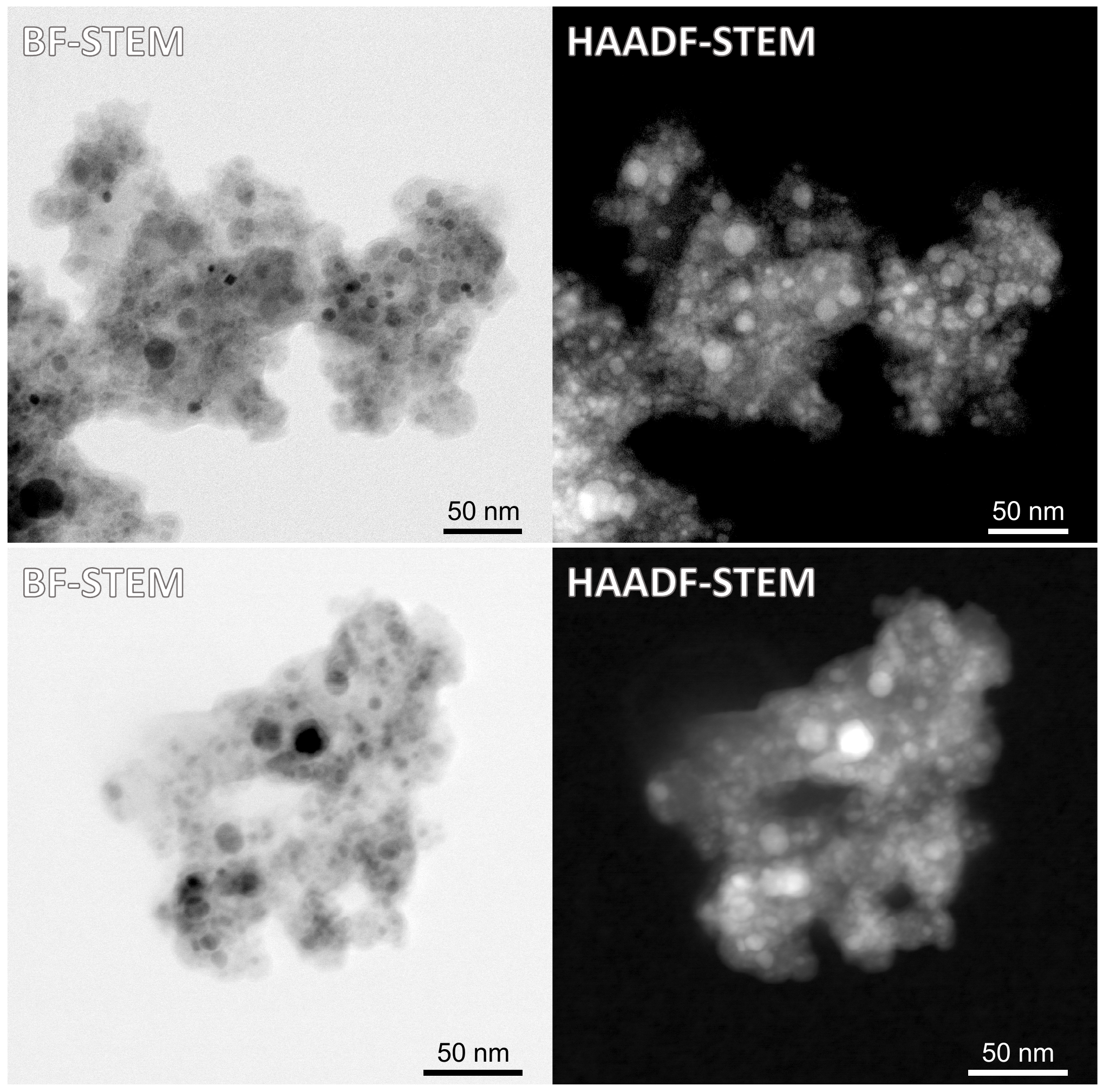}  
        \caption[]{\label{fig:fig_9} %
                BF-STEM and HAADF-STEM images of two typical condensates with GEMS-like textures synthesized in CI-2 and collected from the upper wall.}
\end{figure}
\begin{figure}[hbtp]
        \centering
        \includegraphics[width=\columnwidth]{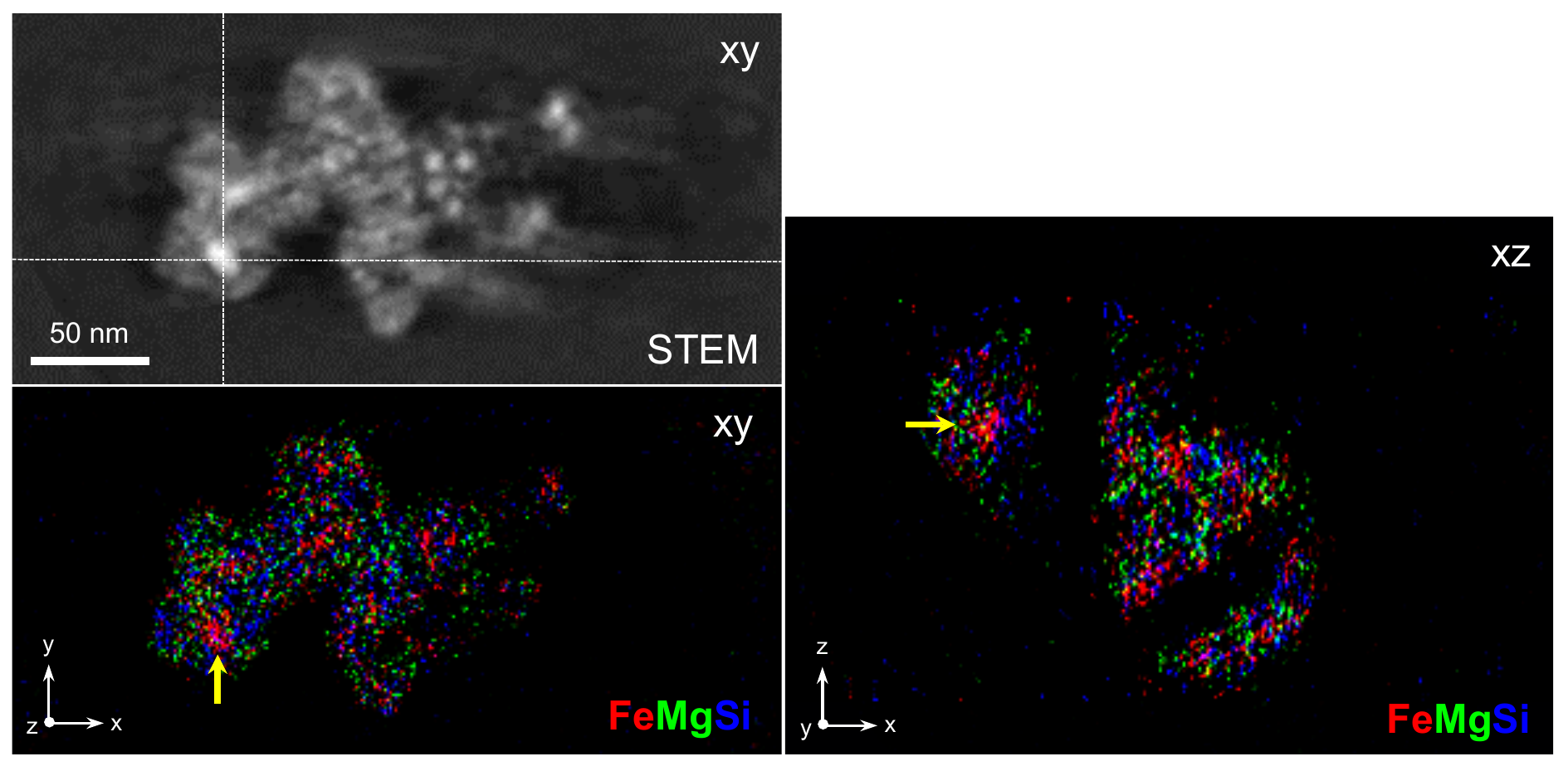}  
        \caption[]{\label{fig:fig_10} %
                Cross-sectional images of DF-STEM and elemental map for x--y and x--z axes to estimate the metallic inclusion of GEMS-like nanomaterial synthesized in CI-2 by TEM tomography. See also online movie.}
\end{figure}

\section{Discussion} \label{sec:Discussion}
\subsection{Condensation process}
Fine particles of \textasciitilde10--100 nm are gas condensates composed of amorphous silicates in the MS experiments (Fig. 5)} and those embedded with nanoparticles of FeNi or Fe$_3$Si in the CI experiments {(Figs. 8 and 10)}. No crystalline silicate was detected among the fine particles.
 
{Spherical grains with clear grain boundaries, which were generally observed for grains on the lower wall of the chamber, have chemically separated textures (Figs. 8 and A.3). These characteristics indicate that they have formed via nucleation of metastable liquid melts.

The amorphous silicate grains {with} Si-rich and Si-poor (Mg-rich) regions {are known to form due to} liquid immiscibility in the systems of MgO--SiO$_2$ \citep{RN2561} and MgO--FeO--SiO$_2$ \citep{RN2574}. {In our experiment, melt droplets seem to} nucleate and grow from vapor as a metastable state {as seen in previous experiments \citep[e.g.,][]{RN2575,RN2563,RN2579}, and} then separate into two liquids {phases} above the glass transition temperature of \textasciitilde800 K \citep{RN2530}. {These melt grains finally solidified as spherical amorphous grains with two distinct chemical compositions.} Silicate melt condensation can occur prior to the condensation of metallic Fe because the temperature of homogeneous nucleation of metallic Fe is lower than that of silicate \citep{RN2570}. The Fe-bearing nanoparticles (Fe$_3$Si and FeNi) inside the amorphous silicate grains observed by the STEM tomography {(Fig. 10; see also online movie)} were likely formed by surface-tension-driven migration of metallic nanoparticles from the surfaces of amorphous silicate grains toward the inside {\citep{RN2570}}.

{The irregularly shaped amorphous silicates (Figs. 3a and 7a, b, d, and e) are aggregates of \textasciitilde10 nm sized particles. The unclear grain boundaries and rough grain surfaces indicate either that (1) irregular shaped small particles condensed from gases and coagulated or (2) very small spherical grains as seen in Fig. 3b condensed and then surface atoms diffused to modify grain shapes and blur grain boundaries to form irregular shapes by the heat of plasma flame. Similar irregularly shaped nanoparticles were observed by condensation experiments of amorphous Si nanoparticles in high quenching gas \citep{RN2580}. The high cooling rate may be a cause of the irregular shaped grains because the quenching rate of the grains at the upper wall of the chamber is higher than that at the lower wall. However, the STEM elemental maps (Fig. 8) show that both of the spherical and irregular shaped grains have Si-rich cores and Mg-rich rims. Thermodynamic calculations of equilibrium condensation predict that Mg-silicates such as forsterite condense at high temperatures and Mg content in silicates decreases at lower temperatures \citep{RN2578}. The exact temperatures of condensation of metastable amorphous silicates are different from those of crystalline silicates but the order of condensation should be similar. Thus, these Si-rich core and Mg-rich rim structures are unlikely to form by condensation of amorphous grains. The difference of the grain shapes collected at the upper and lower walls of the chamber is explained by the difference in the heating after condensation in addition to the quenching rate. The upper wall is only \textasciitilde200 mm from the plasma torch (Figs. 1a and 1b). The condensed solid particles attached to the upper wall were heated at higher temperatures than those at the lower wall and might have been sintered slightly to form ambiguous grain boundaries (Figs. 3a and 7). In experiment CI-3, the samples from both of the upper and lower walls of the chamber are irregularly shaped. This is because the radial plasma gas flow forms a long plasma flame (Fig. 1a) and heated the lower wall of the chamber more than a tangential plasma gas flow.

As shown in Table 1, more than 65\% of the products were collected from the upper {wall} in the MS experiments using tangential gas flow (MS-1--5), while only \textasciitilde25\% collected on the {lower wall} in MS-6, where the radial gas flow was applied. This difference indicates that the starting materials remained longer in the radial plasma flame than in the tangential gas flame and that the main grain formation occurred later in the radial gas flow, as shown in Fig. 1b.

The condensation process was discussed in a model of homogeneous nucleation and growth from vapor \citep{RN2087,RN2570}. In this model, a dimensionless parameter, $\Lambda$, is defined as
\begin{equation} \label{eq:eps}
        \Lambda=\nu\tau/(h/{k_{B}}{T_{e}}-1)
,\end{equation}
where $\nu$, $\tau$, $h$, $kB$, and $T_{e}$ are the gas collision frequency, cooling timescale, evaporation latent heat, Boltzmann constant, and equilibrium condensation temperature, respectively. The collision frequency is given by $\nu=4\pi{a_0}^2{\alpha_S}n_V{(k_BN_AT_e/2)}^{1/2}$, where $a_0$ is the radius of vapor molecules, $\alpha_S$ is the sticking probability, $N_A$, is Avogadro constant, $M$ is molecular weight, and $\nu_V$ is the number density of vapor molecules $({\nu_V}={PX_{ITP}}/{k_BT})$ calculated from the total pressure, $P$, of 7 $\times$ $10^4$ Pa and the abundance of condensate, $X_{ITP}$, of 4 $\times$ $10^{-4}$ by adopting the Si/(Ar + He) ratio using the feeding rate of 100 mg/min. The value of $\Lambda$ in the present ITP experiments, $\Lambda$$_{ITP}$, is estimated to be \textasciitilde4 $\times$ $10^3$, which is comparable to that estimated for the 30 kW ITP system by \citet{RN2570}, at $10^3$--$10^4$. In the calculation, we assume the key species as SiO, a sticking probability of 1, a cooling timescale $(\tau_{ITP})$ of \textasciitilde2.5 $\times$ $10^1$ s, equilibrium temperature of 2,500 K estimated using HSC Chemistry commercial software (Outotec Research Oy.), latent heat of condensation of 31,700 K, and 2.6 $\times$ $10^{-10}$ m radius of vapor molecules \citep{RN1581,RN2571}.
\subsection{Grain size}
\citet{RN1581} proposed that the grain size, $d$, is proportional to $\lambda^{5/6}$. The feeding rate of the starting material, $f$, is proportional to $X_{ITP}$; then, we obtained the following relation:
\begin{equation} \label{eq:eps}
        d \propto (Pf\tau)^{5/6}
.\end{equation}
From this model, the grain diameter of silicate, represented by MgSiO$_3$ \citep{RN1581}, is estimated as \textasciitilde20 nm assuming the experimental conditions of the CI experiments (Table 1). This is consistent with the size of amorphous silicate grains obtained in the CI experiments in magnitude order.

Based on the nucleation and growth model, the grain size, $d$, is almost proportional to the reactor pressure, $P$, feeding rate, $f$, and cooling timescale, $\tau$, as shown by Eq. (2). If we assume that $\tau$ is the same in the tangential flow without the slit gas, the value of $Pf^{5/6}$ increases from MS-4 to MS-3 to MS-2 to MS-5 (Table 1), which is consistent with the difference in median grain size of products collected from the {lower wall (Fig. 4a)}. In addition, the grain size in MS-1, in which an additional O$_2$ gas was introduced, is smaller than those in MS-2--5. The condensates in the radial gas flow (MS-6) are larger than those in MS-2--5. These results indicate that the cooling timescale became smaller in MS-2 due to the additional gas flow, and larger in MS-6 due to the radial gas flow, respectively. In addition, the grain size in the upper {wall} of MS-1--5 is smaller than that in the {lower wall (Figs. 4b and c)}. This suggests that the cooling rate in the upper {wall} is larger ($\tau$ is smaller) than that in the {lower wall} under the tangential gas flow conditions. The coexistence of fine grains (\textasciitilde10 nm) with coarse grains in the {lower wall} suggests that heterogeneous condensation conditions in the reactor and that grains of different sizes were finally mixed together. In the CI experiments, no significant difference in particle size was noted in CI-1--3. The value of $Pf^{5/6}$ is much smaller in CI-3 than that in CI-1 and CI-2 (Table 1), although $\tau$ in the radial gas flow is larger than that in the tangential gas flow, as observed in the MS experiments. Although the values of $\tau$ could not be estimated quantitatively in this study, similar grain size was obtained by compensation of $Pf$ and $\tau$ in the CI experiments.
\subsection{Comparison with GEMS-like nanoparticles synthesized in higher power ITP}
\citet{RN2570} also produced spherical amorphous silicates embedded with multiple metal sub-grains similar to GEMS. Those grains are, however, constituted only part of the condensates; the remaining grains were amorphous silicate grains, either without inclusions or with a single metal core. However, in the present study, the condensates from the upper {wall} in CI-2 are composed mainly of irregularly shaped amorphous silicate with multiple metallic inclusions similar to GEMS {(Fig. 9)}. These differences could have been caused by different temperature gradients and locations of the sample collection in the ITP systems. The higher power ITP system of 30 kW used in \citet{RN2570} provided a larger volume of high-temperature gases than the present ITP system of 6 kW. The condensates were collected >1 m from the plasma torch in the 30 kW ITP system. {The small fraction of the amorphous silicates embedded with multiple metal sub-grains in \citet{RN2570} indicates that amorphous silicate grains with and without multiple metal inclusions formed in different locations and different conditions in the 30 kW ITP system and were collected together.
	
\citet{RN2570} argued that} the spherical shape of the amorphous silicates embedded with multiple metallic inclusions synthesized in the 30 kW ITP system {are} formed by either (1) heterogeneous condensation of metals on the silicate melt surface followed by incorporation of metals into the silicate melt before solidification or by (2) coalescence of silicate melt grains before solidification, each of which had a single metallic core.

On the contrary, as discussed in {Sect. 4.1}, our study showed that the condensed solid particles attached to the {upper wall} were heated because of the short distance from the plasma torch and were sintered slightly to form ambiguous grain boundaries without crystallization. The irregularly shaped amorphous silicates with multiple metallic inclusions collected from the upper {wall} in CI-2 are explained by solidification of small amorphous silicate grains of less than a few tens of nanometers with a metallic inclusion (core-shell grains) followed by coagulation and {annealing} of those grains.
\subsection{Implication for GEMS formation}
The textures of multiple metallic nanoparticles embedded in irregularly shaped amorphous silicate grains of \textasciitilde10 to 100 nm produced in the CI experiments are similar to those of GEMS in CP-IDPs \citep{RN306,RN305,RN304} and chondritic porous micrometeorites (CP-MMs) \citep{RN2573,RN2560,RN2559,RN2531,RN2576}. Mid-IR spectra of condensates in the CI experiments well reproduce peak positions of GEMS and circumstellar amorphous silicates \citep[e.g.,][]{RN2402}. In addition, the signatures of the liquid immiscibility (SiO$_2$-rich and poor compositions in amorphous silicates) can be recognized in GEMS \citep[e.g., STEM-EDS images in Fig. 1 of][]{RN304} and possibly the IR spectrum of GEMS \citep{RN2402}. As discussed {in Sect. 4.1}, this liquid separation texture was likely formed by condensation as a melt state at a} relatively high temperature above the glass transition temperature.

The size of the amorphous silicate spheres in the CI experiments was somewhat smaller than that of the GEMS grains. However, some GEMS grains appear to be aggregates of sub-grains, based on their TEM images \citep{RN304}. The texture of the irregularly shaped amorphous silicate grains with embedded multiple nanoparticles produced in CI-2 (Fig. 9) is very similar to GEMS except for the presence of Fe$_3$Si and the absence of FeS. 
The Fe$_3$Si nanoinclusions observed in the CI experiments are not observed in GEMS. This indicates that the redox conditions of the GEMS formation are more oxidized than those in the CI experiments. In addition, we cannot discuss Fe sulfide nanoinclusions observed in GEMS because the experiments were performed in the S-free system. Sulfidation {may} occur at low temperatures simultaneously with {annealing} of amorphous silicates.

Although the discussion using $\Lambda$ values can apply only to the grain size of condensed particles and not to textures or phases of the condensates, we compared the condensation process in the ITP system with GEMS formation, which occurred in the primordial Solar System or the circumstellar environments of evolved stars. The condensation parameter of $\Lambda$$_{ITP}$ of \textasciitilde~4 $\times$ 10$^3$ is comparable to those around Type II-P supernovae (SNe), outflow of asymptotic giant branch stars, and bow shock regions in protoplanetary disks but is not similar to those of slow-cooling gas of protoplanetary disks at 0.1 AU (cooling timescale of \textasciitilde3 $\times$ 10$^8$ s) or around Type IIb SNe, as proposed by \citet{RN2570}. It should be noted that the values of $\Lambda$$_{ITP}$ are essentially the same among the experiments in this study, which indicates the possibility of GEMS-like texture forming in the other environments. If condensation of amorphous silicate of tens of nanometers with a metallic core followed by coagulation and slight {annealing} discussed in the previous section is the key process in forming GEMS, lower $\Lambda$ environments such as Type IIb SNe are possible candidates for GEMS formation as well because of the relation $\Lambda$ $\propto$ $d^{5/6}$. In that case, such small grains need to coagulate prior to destruction by shock waves from SNe.

\section{Conclusion}     \label{sec:Conclusion}
We performed vaporization and condensation experiments in the Mg--Si--O and Si--Mg--Fe--Na--Al--Ca--Ni--O systems under nonequilibrium conditions in the 6 kW ITP system.
 
Vaporization and condensation conditions (condensation location, cooling rate, reactor pressure, and plasma flame pattern) in the ITP system were examined by experiments in the Mg--Si--O system (MS experiments). The condensates collected at the upper {wall} are smaller than those at the {lower wall} in most MS experiments owing to the difference in the paths of the gas flow, which reached the different locations in the chamber and changed the temperature and density gradient of the gas molecules. The grain size distributions in the MS experiments clearly depended on the plasma conditions and are well explained by nucleation and growth theory \citep{RN2087,RN1581}.

{In the} condensation experiments of gases of the CI chondritic composition in the system of Si--Mg--Fe--Na--Al--Ca--Ni--O, amorphous silicates of a few tens of nanometers with multiple metallic sub-grains {were} collected at the upper {wall} in CI-2, {and their textures} appear very similar to the characteristic textures of GEMS, except for the presence of Fe$_3$Si grains instead of sulfide grains. These GEMS-like materials are formed via condensation of small {liquid} grains of less than a few tens of nanometers with metal cores followed by the aggregation of condensed grains and a slight {annealing} without crystallization by heat from the hot regions. {The Fe content in amorphous silicate grains in the CI experiments increased with a decrease in the feeding rate of the starting materials, likely because the redox conditions were affected by the background oxygen in the chamber.

We compared the condensation conditions in the ITP system with circumstellar environments using the nucleation and growth theory. The formation of amorphous silicate grains of a few tens of nanometers in size in the protoplanetary disk requires local thermal events such as chondrule formation. Circumstellar environments, such as Type II-P SNe, Type IIb SNe, and the outflows of AGB, are also possible environments. A slightly oxidized condition for preventing the formation of silicide and the sulfidation of metals at low temperatures may be required to more effectively reproduce the GEMS mineralogy.

The formation processes for reproducing GEMS textures proposed in the present study are (1) the condensation of silicate melts followed by a heterogeneous condensation of metal onto them, (2) surface-tension-driven migration of metallic nanoparticles from the surfaces to the inside of silicate melts, (3) solidification and coagulation of the spherical amorphous silicate grains, and (4) slight {annealing and diffusion of the surface atoms to form} irregularly shaped amorphous silicate with multiple nanoinclusions. Although the conditions for the {annealing} of amorphous silicates to form irregular grain shapes are not constrained from the present study, {the heating} of amorphous silicates at temperatures lower than their crystallization temperature does not conflict with later sulfidation after the condensation of amorphous silicate and nanometal or isotopic homogeneity between the GEMS grains.
\begin{acknowledgements} 
        This study was financially supported by the JSPS KAKENHI Grant Numbers 15H05695 and 20H00205 (Tsuchiyama) and 19H01935 and 19H00712 (Takigawa). A. Tsuchiyama was also supported by Chinese Academy of Sciences International Fellowship for Visiting Scientists Grant No.2019VCA0004.
        
\end{acknowledgements}
\bibliographystyle{aa-note} 
\bibliography{AA_KTH_rev04}

\begin{appendix} 
        \section{Unvaporized and melted residues}
Figure A.1 shows backscattered electron (BSE) images of the run products collected from the upper wall in MS-1, MS-5, and MS-6. Matrix regions in Fig. A.1 consist of condensates of <100 nm in size. Large spherical and nonspherical grains of about one to a few tens of micrometers in size are embedded in the condensates (matrix). The large grains observed in SEM are not condensates but unvaporized or unmelted residues of the starting materials because the particles cannot grow to a micron size in typical ITP systems at an extremely high cooling rate \citep{RN2565}. Also, the grain size estimated in Sect. 4.2 showed that condensates should be <100 nm in size. Large grains of periclase and quartz (>5 $\mu$m) are vaporization residues because they have similar shape, texture, and size as those of the starting reagent grains. Micron sized spherical grains are clearly larger than the condensates and the grain shape indicates that they are produced by melting of the starting materials. Their compositions show variety in MgO/SiO$_2$ ratios and are sometimes similar to forsterite (Mg$_2$SiO$_4$) and enstatite (MgSiO$_3$). Forsterite and enstatite were crystallized from the melts during cooling. The proportion of these evaporation residues (both unevaporated and melted) increases from MS-1 to MS-6 (Fig. A.1) and is higher in the quenching part than that in the upper and lower walls of the chamber.

The matrices are composed of fine grains (<100 nm). A similar fine grain size was observed in the condensation products created by rapid quenching of gases in previous ITP experiments \citep[e.g.,][]{RN2524}; thus, those in the present study are condensation products. SEM-EDS analysis showed that the chemical composition of the matrix is uniform and depleted in Mg compared with that in the bulk starting materials (Mg/Si = 1), which is explained by incomplete vaporization of Mg-rich materials. The ratio of Mg/Si in the bulk nanoparticles increases from \textasciitilde0.7 to \textasciitilde1 from MS-1 to MS-6.

Figures A.1c and A.1d compare the run products at high reactor pressures (\textasciitilde70 kPa) with different plasma flame patterns of tangential and radial for MS-5 and MS-6, respectively. The run products are composed mostly of fine condensates, particularly in MS-6. Most evaporation residues are partially melted and recrystallized sub-rounded MgO particles and Mg-rich amorphous silicate spheres, suggesting that the SiO$_2$ was almost completely evaporated under these conditions. The Mg/Si ratios of the fine matrix are \textasciitilde1 and \textasciitilde0.8 in MS-6 and MS-5, respectively, suggesting that the vaporization efficiency was higher by the radial plasma flame (MS-6) than the tangential flame (MS-5).

Figure A.2 shows the XRD patterns of the starting material and run products in MS experiments. All XRD patterns of run products have a 2$\theta$ halo at about 20$^{\circ}$--40$^{\circ}$ by amorphous silicates, which were contributed mainly from condensates and to a lesser extent from quenched silicate melts. Crystalline peaks were assigned to periclase and quartz (evaporation residues) and forsterite and protoenstatite (MgSiO$_3$) as crystallization products from melts by quenching, as shown in the SEM observations (Fig. A.1). The small intensities of quartz peaks in MS-1 in contrast to the abundant SiO$_2$ particles in the SEM images (Fig. A.1) suggest that most of the quartz particles were melted and became amorphous. The amount of evaporation residues compared with those of amorphous silicates are highest in the quenching part (Fig. A.2), and that of amorphous silicates increases from MS-1 to MS-6.

The incomplete vaporization is mainly attributed to the low power of the present ITP system (6 kW) compared with the previous experiments at 30 kW \citep{RN2570,RN2524}. In particular, MgO was more difficult to vaporize than SiO$_2$ because of the high melting/boiling temperatures (2,852/3,600 $^{\circ}$C~for MgO and 1,650/2,230 $^{\circ}$C~for SiO$_2$, \citep{RN2562}). The crystallinity of unvaporized and melted residues could reflect the viscosity of the melts. Melted MgO grains recrystallized during cooling because of the very low viscosity of the MgO melt. In contrast, the SiO$_2$ melt hardly crystallized by rapid quenching because of its very high viscosity; thus, we obtained many amorphous silica residues (Figs. 2 and A.2). If grains of different materials were in contact with each other, eutectic melting would have occurred at lower temperatures to form multi-component melts (e.g., MgO-SiO$_2$ melt). During cooling, forsterite and enstatite crystallized from such melts and the residual melts became amorphous.
\begin{figure}[hbtp]
        \centering
        \includegraphics[width=\columnwidth]{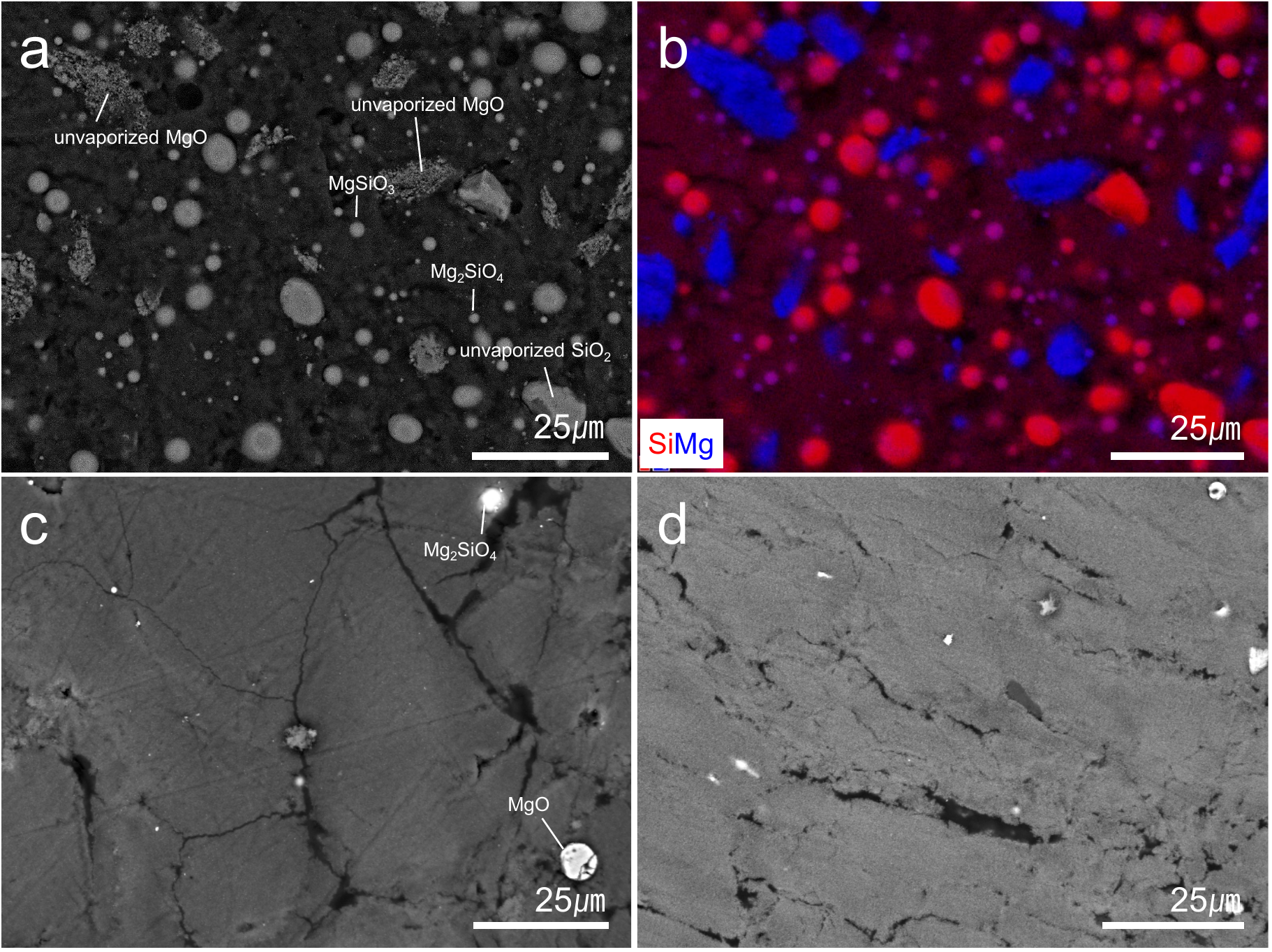}  
        \caption[]{\label{fig:fig_A_1} %
                BSE image (a) and elemental map (b) of a polished section of run products in MS-1.  BSE images of MS-5 (c) and MS-6 (d). These samples were collected from the upper wall.
        }
\end{figure}

\begin{figure*}[hbtp]
        \centering
        \includegraphics[width=\hsize]{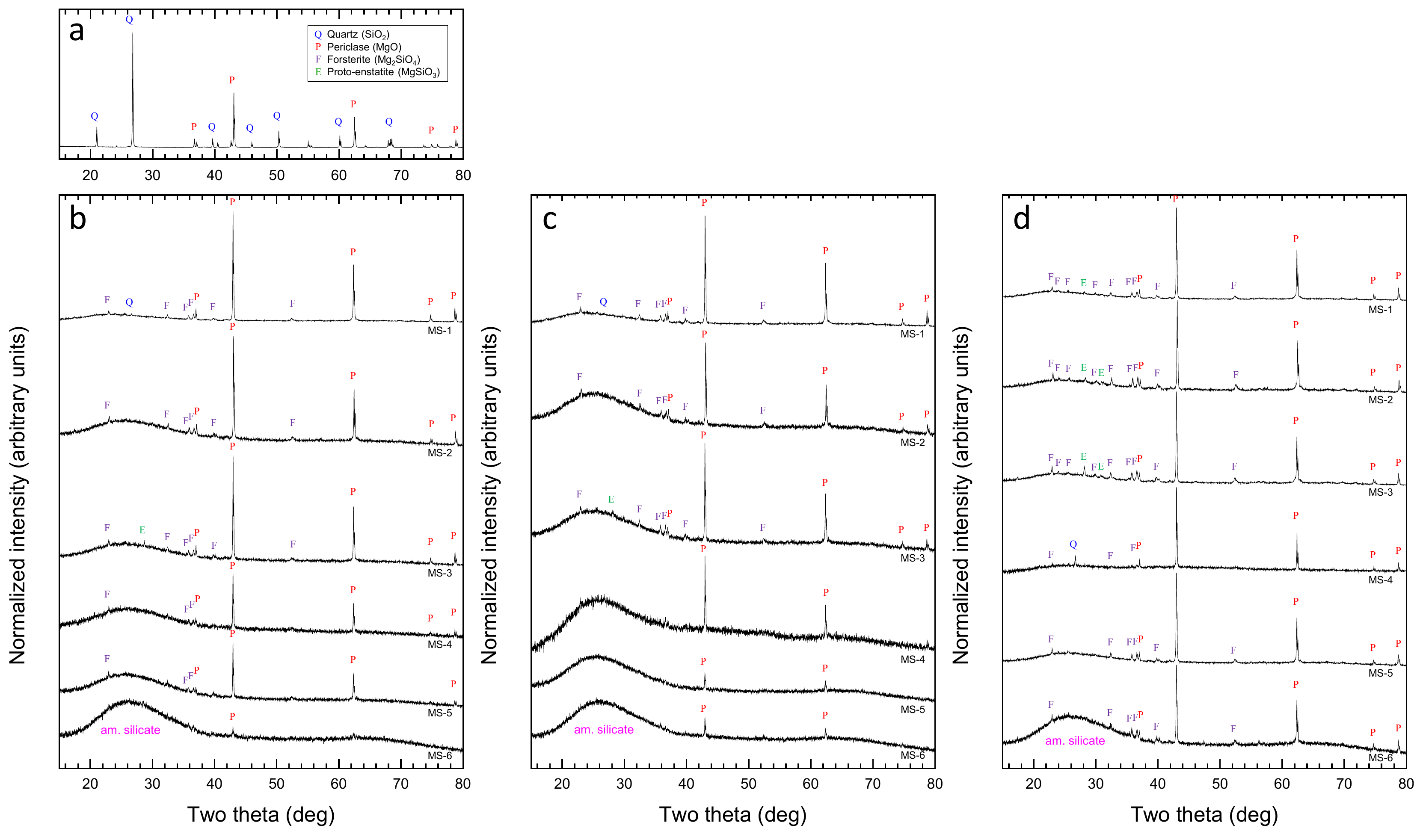}  
        \caption[]{\label{fig:fig_A_2} %
                XRD patterns of starting material (a) and run products collected at the upper wall (b), lower wall (c), and  quenching part in MS-1--6 (d) (MgO--SiO$_2$ system).
        }
\end{figure*}

\begin{figure}[hbtp]
        \centering
        \includegraphics[width=\columnwidth]{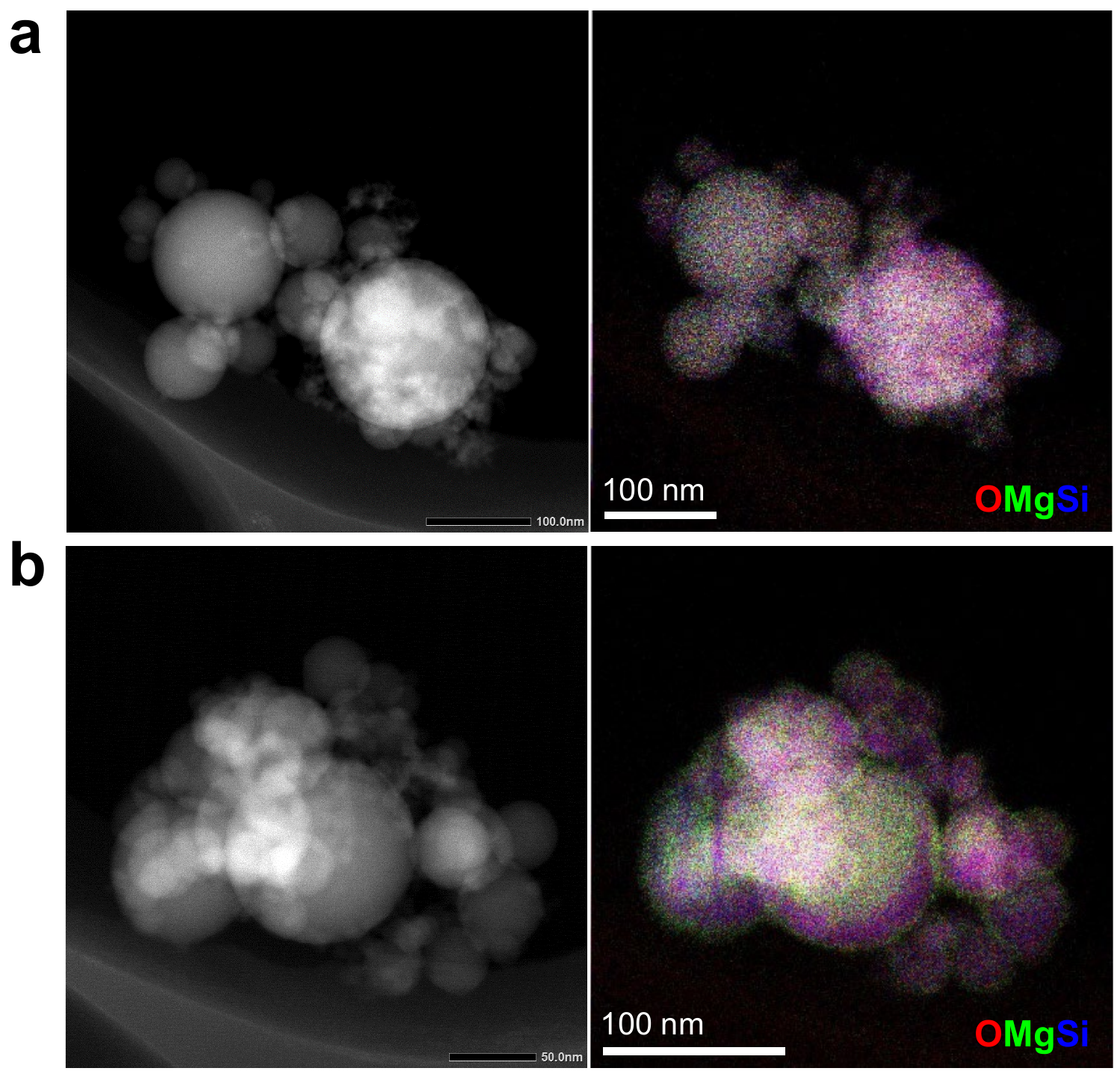}  
        \caption[]{\label{fig:fig_A_3} %
                DF-STEM images and STEM-EDS maps of run products synthesized in MS-6 collected from the upper wall (a) and lower wall (b).
        }
\end{figure}

        \section{Optimal conditions for evaporation in the experiments}
We examined the (1) presence of slit gas, (2) reproducibility, (3) reactor pressure, and (4) plasma flame patterns in the MS experiments to obtain the optimal conditions for minimizing the amount of evaporation residues (Table 1).

The experimental conditions of MS-1 and MS-2 were the same except for the presence of the O$_2$ slit gas (Table 1). The slit gas provides a faster quenching rate of the vapor prior to thermal expansion. The intensity ratio between the amorphous halo and the periclase peaks in the XRD patterns is larger in MS-2 than in MS-1 (Figs. A.2b--d), indicating that application of the slit gas negatively affected the vaporization of the starting material. Reproducibility of the experiments was confirmed by MS-2 and MS-3, in which the experiments were performed under essentially the same conditions, and the characteristics of the obtained condensates were very similar.
 
The reactor pressure was set at 30, 50, and 70 kPa in MS-3, MS-4, and MS-5, respectively (Table 1). The ratio of amorphous silicate and periclase increased with pressure (Figs. A.2b--d). A higher reactor pressure lowered the flow velocity and shortened the plasma flame. A slow plasma velocity can lead to a long residence time of the starting material at the high-temperature region, thus improving the vaporization efficiency.

The run products of MS-6, which adopted the radial flow of the plasma-forming gas, have strong amorphous halos with very weak periclase peaks in both the upper and lower walls (Figs. A.2--d). The tangential flow facilitated rapid quenching due to the widespread flame (e.g., MS-5), whereas the radial flow provided a longer and narrower plasma flame owing to the vertical gas injection (Fig. 1b). This improved the residence duration of the starting materials at the high-temperature region of the plasma flame and yielded high vaporization efficiency.
  
The current results indicate that reactor pressure was the most influential parameter for the evaporation efficiency followed by the flow pattern of the plasma-forming gas. The optimal condition for minimizing the vaporization residue was high pressure with the radial flame, such as that used in MS-6. We conducted the CI experiments based on the results and found a similar optimal condition of 70 kPa with the radial flame in CI-3. The gas flow pattern was not restricted to the radial flow because the amount of vaporization residue was not as high at high pressure of \textasciitilde70 kPa even in the tangential flow (MS-5 in Fig. A.2 and CI-2 in Fig. 5b).     

\end{appendix}

\end{document}